\documentclass[11pt,onecolumn,floatfix,notitlepage]{article}
\usepackage[margin=1.0in]{geometry}
\usepackage{graphicx}
\usepackage{braket}
\usepackage[justification=justified,
   format=plain]{caption} 
\usepackage{subcaption}
\usepackage{bbm} 
\usepackage{amsmath}
\usepackage{appendix}
\usepackage{authblk}
\usepackage[math]{cellspace}
\usepackage{xcolor}
\usepackage{multirow}
\usepackage[english]{babel}

\title{Born Machines for Periodic and Open XY Quantum Spin Chains}

\date{\today}

\author[1]{Abigail McClain Gomez}
\author[1]{Susanne F. Yelin} 
\author[1,2]{Khadijeh Najafi}
\affil[1]{Department of Physics, Harvard University, Cambridge, Massachusetts 02138, USA}
\affil[2]{IBM Quantum, IBM T.J. Watson Research Center, Yorktown Heights, NY 10598 USA}
\begin{document}

    \maketitle
    \begin{abstract}
    Quantum phase transitions are ubiquitous in quantum many body systems. The quantum fluctuations that occur at very low temperatures are known to be responsible for driving the system across different phases as a function of an external control parameter. The XY Hamiltonian with a transverse field is a basic model that manifests two distinct quantum phase transitions, including spontaneous $Z_2$ symmetry breaking from an ordered to a disordered state. While programmable quantum devices have shown great success in investigating the various exotic quantum phases of matter, in parallel, the quest for harnessing machine learning tools in learning quantum phases of matter is ongoing. In this paper, we present a numerical study of the power of a quantum-inspired generative model known as the Born machine in learning quantum phases of matter. Data obtained from the system under open and periodic boundary conditions is considered. Our results indicate that a Born machine based on matrix product states can successfully capture the quantum state across various phases of the XY Hamiltonian and close to a critical point, despite the existence of long-range correlations. We further impose boundary conditions on the Born machine and show that matching the boundary condition of the Born machine and that of the training data improves performance when limited data is available and a small bond dimension is employed. 
\end{abstract}

\section{Introduction}
\label{1Introduction}

Quantum many body systems have attracted the attention of physicists for many years. These systems are far from being understood by any simple analytical solution; their many body nature demands more elaborate techniques that involve statistical approaches. Moreover, the competition between various degrees of freedom (such as motion, charge, and spin) is the origin of many exotic phases of matter, ranging from quantum magnets to superconducting and topological phases. Recent developments in building coherent programmable quantum devices have provided a compelling tool to probe the fundamental properties of the quantum matter, and the quest for building and controlling even larger devices is an ongoing process \cite{Misha1,Ebadi2021}. However, in parallel with the growth in system size, the number of measurements required for the full reconstruction of quantum state scales exponentially, and the noisy nature of the data becomes more pronounced. 

Remarkably, machine learning (ML) tools have already shown great promise in extracting effective information from large amounts of noisy data \cite{Iris_QCNN,huang_QML_2021}. One category of machine learning applications that are inherently quantum is the task of understanding or characterizing physical systems that exhibit many body quantum effects, such as correlated electronic structure or phase transitions from disorder to order in spin systems. Due to the complex behavior of these systems, mean-field or semi-classical approaches are insufficient, yet the naive approach of working in the entire quantum state space would scale exponentially. Concretely, the exponential scaling of such systems makes any scalable simulation on classical devices unfeasible. Quantum simulations on quantum computers could lead to efficient simulations for many body quantum systems or, more generally, any sparse quantum Hamiltonian. However, it is not simple to extract meaningful knowledge of a physical observable from a finite set of local measurements on a noisy processor, which are prone to error and very expensive to query. 

\begin{figure}[htp]
    \centering
    \begin{subfigure}[b]{0.49\textwidth}
        \centering
        \captionsetup{justification=raggedright,singlelinecheck=false}
        \includegraphics[scale = 0.12]{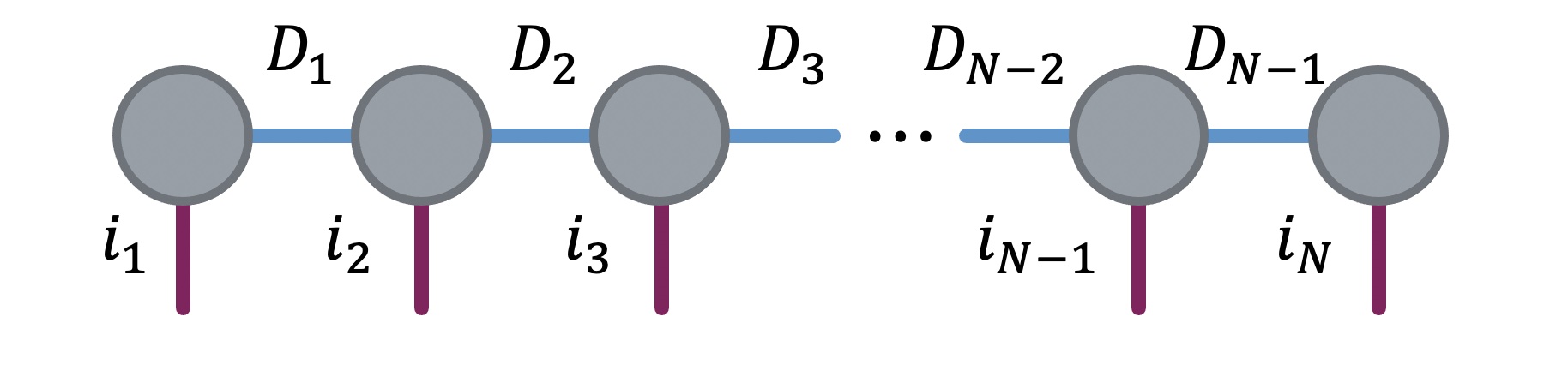}
        \caption{}
        \label{fig:born_machine_open}
    \end{subfigure}
    \hfill
    \begin{subfigure}[b]{0.49\textwidth}
        \centering
        \captionsetup{justification=raggedright,singlelinecheck=false}
        \includegraphics[scale = 0.12]{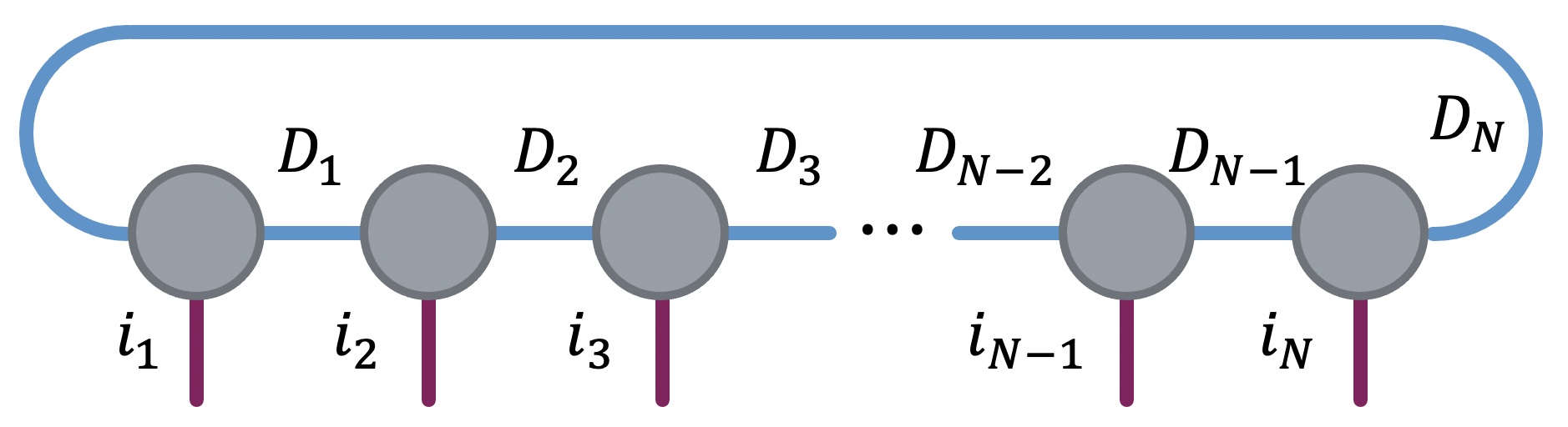}
        \caption{}
        \label{fig:born_machine_periodic}
    \end{subfigure}
    \caption{(a) Open Born machine architecture for a chain of $N$ atoms. Here, $\{i_1, ... i_N\}$ denote the spin indices (which all have size 2 for a two-level model of each atom) and $\{D_1, ... D_{N-1}\}$ denote the bond dimensions (which all have size $D$ for our architectures). (b) Periodic Born machine architecture for a chain of $N$ atoms. Note that there are $N$ bond dimensions in this architecture.}
    \label{fig:born_machine_arch}
\end{figure}

While finding efficient machine learning-based algorithms for quantum state tomography remains an important question, another intriguing possibility is the direct characterization of quantum systems through the measurement of many-body correlation functions. In particular, the behaviour of non-equilibrium quantum systems undergoing a quantum phase transition (QPT) has attracted much attention. Such transitions are driven by quantum fluctuations at zero temperature and can be identified by the divergence of the correlation length, an encompassing universal behaviour near the critical point \cite{Polkovnikov,Misha2}. In this paper, we address these questions by leveraging advanced ML algorithms that enable quantum simulation. Specifically, we explore the task of learning various phases of the one dimensional XY Hamiltonian with open and periodic boundary conditions using the Born machine (BM). We further introduce the Periodic Born machine (see Fig. \ref{fig:born_machine_arch}) and show that the architecture of the Born machine is sensitive to the boundary condition imposed on the training data.
\begin{figure} [htb] 
\centering
\includegraphics[width=0.40\textwidth]{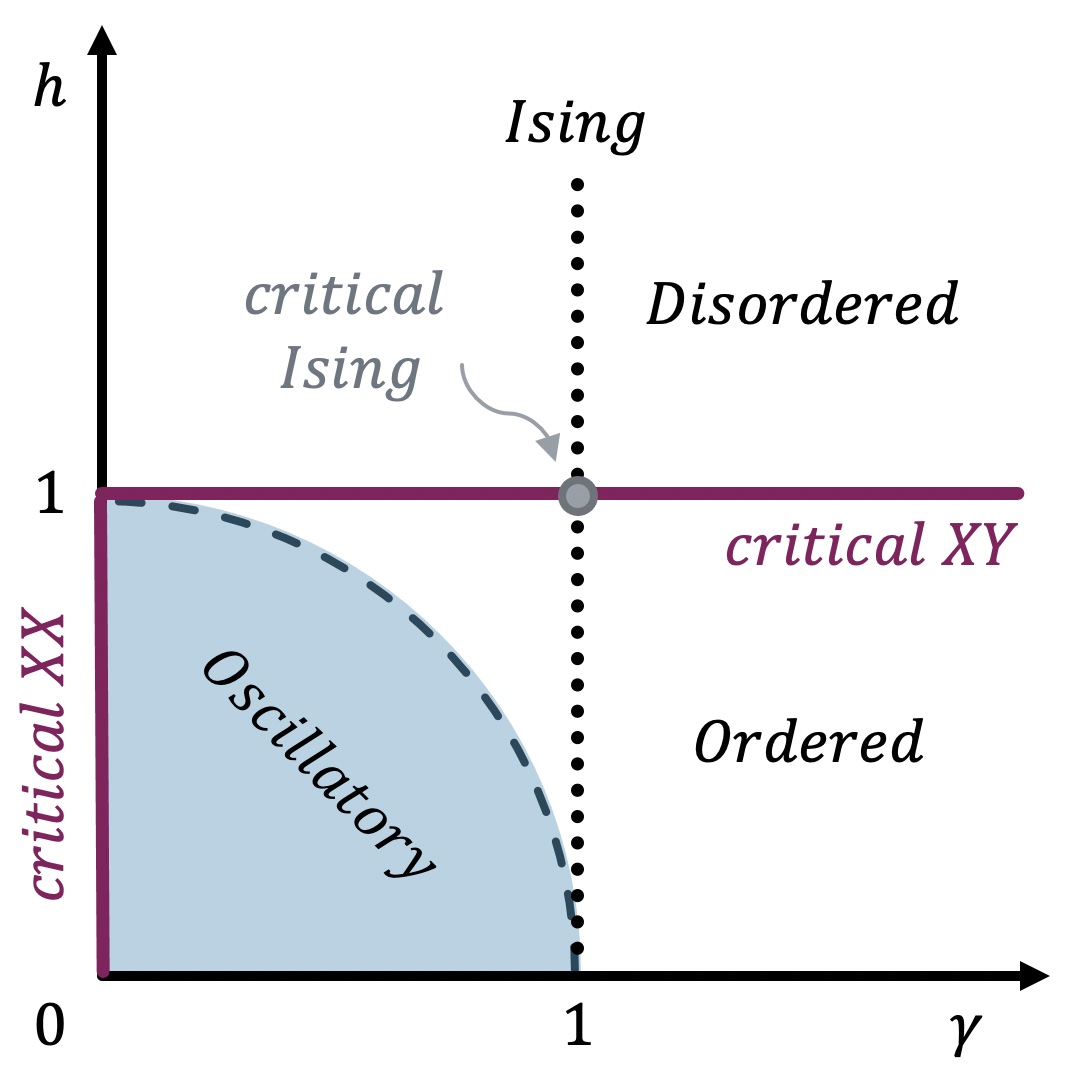}
\caption{Different regions in the phase diagram of the quantum XY chain depicted for positive values of $\gamma$ and $h$. The model is critical for $h=1$ as well as for $h<1, \gamma = 0$, which respectively correspond to the critical XY and XX chains. The model manifests a quantum phase transition from disorder ($h>1$) to ordered phase ($h<1$). The point $(\gamma=1 , h=1)$ corresponds to the critical point of the transverse field Ising model, and on the line $\gamma^2+h^2=1$, the ground state can be factorized into a product of single spin states.}
\label{Phase_XY}
\end{figure}
We start with the Hamiltonian of the XY model in the transverse magnetic field, described by:
\begin{eqnarray}\label{H_XY}\
\mathbf{H}=-J\sum_{i = 1}^{N-1}\Big{(}\frac{1+\gamma}{4}\sigma_i^x\sigma_{i+1}^x+\frac{1-\gamma}{4}\sigma_i^y\sigma_{i+1}^y\Big{)}-\frac{h}{2}\sum_{i = 1}^N\sigma_i^z
\end{eqnarray} 
where $J$ is the spin-spin coupling, $\gamma$ is an asymmetric constant capturing the relative strength of interaction along $x$ and $y$, and $h$ is the external magnetic field coupled to spins along the $z$-axis. Despite the fact that this Hamiltonian can be mapped to the free fermionic model via the Jordan-Wigner transformation, it manifests a rich phase diagram (see Fig. \ref{Phase_XY}) \cite{Franchini}. The phase diagram involves two distinct quantum phase transitions; namely, the critical XX line corresponding to $\gamma=0$ with $h\leq1$ (belonging to $U_1$ symmetry) and the critical XY line corresponding to $h=1$ (belonging to the $Z_2$ symmetry sector). The latter transition occurring at $(|h|=J=1)$ is an Ising like transition, in that it exhibits spontaneous breaking of $Z_2$ symmetry. This transition from a doubly degenerate state in the ordered phase for $|h|<1$ to a single ground state ($|h|>1$) is similar to the classical 2D Ising phase transition, with the magnetic field replacing temperature. The XY model is exactly solvable, allowing one to attain analytical solutions for many interesting local and non-local quantum observables and their corresponding time evolutions \cite{Najafi1,Najafi2,Najafi3,Najafi4}. These solutions can serve as valuable bench-marking tools for probing the system close to a critical point \cite{Najafi1}. In this work, rather than calculating the exact solution, we used the density matrix renormalization group (DMRG) for the simplicity of the numerical model. DMRG has shown great success in evaluating many physical observables of interest for quantum spin chains \cite{DMRG1,DMRG2}.

The rest of the paper is organized as follows. In Section~\ref{2.Born Machine}, we first introduce the quantum inspired generative model known as the Born Machine, based on an important category of quantum many body states called matrix product states (MPS) \cite{Wang_MPS,Wang_TTN}. We further introduce the Periodic Born machine and demonstrate its ability to learn the underlying boundary condition of quantum training data. In Section \ref{3.set up}, we present various data from different points of the phase diagram of the XY quantum spin chain obtained using DMRG simulations with ITensor \cite{itensor}. We then examine a successful example of learning a ground state of the XY Hamiltonian with the Born machine. Subsequently, we present our results for different boundary conditions with a discussion of how the loss function and quantum fidelity of the model is affected. In Section~\ref{4.discussion}, we conclude the paper with a brief discussion of our results, an investigation into the conditions that cause the Born machine to fail, and an explanation of how these obstacles can be overcome by generalizing the algorithm to a complex-valued basis-enhanced Born machine. 
\section{Open and periodic Born machines}
\label{2.Born Machine}
A generative model has two distinct objectives: 1) to learn some joint probability distribution from training data, and 2) to subsequently generate new data from the trained model. The Born machine is a kind of generative machine learning model with a quantum-inspired framework. Rather than directly modeling the joint probability distribution, the Born machine models the quantum wave function $\Psi$, or more generally, the probability amplitude distribution \cite{Wang_MPS,Wang_TTN}. To differentiate the actual quantum state $\Psi$ and the estimated quantum state modeled by the Born machine, we denote our trained model with a hat: $\hat{\Psi}$. One can arrive at the probability distribution of the model through the probabilistic interpretation of quantum mechanics, also known as Born's rule:
\begin{eqnarray} \label{eq:Borns Rule}
    \hat{\mathbbm{P}}(\mathbf{v}) = \frac{\lvert \hat{\Psi}(\mathbf{v})\rvert^2}{\sum_{\mathbf{v}\in \mathcal{V}}{\lvert \hat{\Psi}(\mathbf{v})\rvert^2}}
\end{eqnarray}
Here, $\mathcal{V}$ refers to the entire Hilbert space. Training occurs through the minimization of the negative log likelihood (NLL), defined below:
\begin{eqnarray}\label{eq:OG Loss}
    \mathcal{L} = -\frac{1}{\lvert \mathcal{T} \rvert}\sum_{\mathbf{v} \in \mathcal{T}} \ln \hat{\mathbbm{P}}(\mathbf{v})
\end{eqnarray}
In the above expression, $\hat{\mathbbm{P}}(\mathbf{v})$ is the probability distribution of our model, defined in Eq. (\ref{eq:Borns Rule}), and $\mathcal{T}$ represents the set of training data. Minimization of the NLL will maximize the model's probability of configurations $\mathbf{v}$ that appear in the training data. If the probability distribution $\hat{\mathbbm{P}}(\mathbf{v})$ perfectly captures the probability distribution of the training data $\mathbbm{P}(\mathbf{v})$, then the NLL will approach the Shannon entropy of the training data $\mathcal{T}$: $S = -\sum_{\mathbf{v} \in \mathcal{T}} {\mathbbm{P}}(\mathbf{v}) \ln({\mathbbm{P}}(\mathbf{v}))$. 

The matrix product state (MPS) is a natural and efficient way of classically representing the quantum wave function of a one-dimensional chain of atoms: 
\begin{eqnarray}\label{MPS}\
{\hat{\Psi}} &=& \sum\limits_{D_1 = 1}^{D}\cdots \sum\limits_{D_{N-1} = 1}^{D}{ M_{1,i_1}^{D_1}\, M_{2,i_2}^{D_1,D_2}\, \cdots \,M_{N,i_{N}}^{D_{N-1}}} \\
\label{Expanded_Psi}
&=& \sum\limits_{\mathbf{v} = 1}^{2^N} \hat{\Psi}(\mathbf{v})\lvert \mathbf{v}\rangle 
\end{eqnarray} 
The tensors listed in Eq. (\ref{MPS}) can be contracted along each interior bond dimension $D$ to produce a single rank-N tensor, which can be cumbersomely expressed with $2^N$ parameters $\hat{\Psi}(\mathbf{v})$ in the computational basis -- see Eq. (\ref{Expanded_Psi}). Here, $\mathbf{v}$ again represents a single configuration in the computational basis ($\lvert \mathbf{v}\rangle  = \lvert \lambda_1, \lambda_2, \cdots , \lambda_N\rangle, \lambda_j \in \{0, 1\})$.

It is known that the expressive power of matrix product states arises from the amount of entanglement encoded in their bond dimensions, which depends on the particular architecture. Taking advantage of this fact, we use an MPS as the ansatz for the model with one tensor for each atomic site. As shown in Fig. \ref{fig:born_machine_open}, our selected MPS architecture involves an array of rank-3 tensors connected to each other via a virtual bond dimension of size $D$, while the dangling bonds indicated by $i$ are spin indices. The parameters trained by the Born machine are simply the elements of each tensor in the MPS. For simplicity, we use the same bond dimension $D$ between each tensor and select $D$ prior to learning (although in theory each bond dimension could be adaptively learned as well). 

We employ two subtly distinct MPS architectures. The first involves an open boundary condition and is referred to as the Open Born machine. In this MPS architecture, the first and last tensors have only two indices: one spin index and one bond dimension linking the tensor to the rest of the chain. The second MPS architecture involves a periodic boundary condition and is aptly named the Periodic Born machine. In this case, each tensor has three indices: one spin index and two bond dimensions linking the tensor to its neighbors on either side. A bond dimension is added between the first and last tensor to complete the ring.

\section{Learning a one dimensional XY quantum spin chain}
\label{3.set up}

\begin{figure} [htb] 
    \centering
    \includegraphics[width=1.0\textwidth]{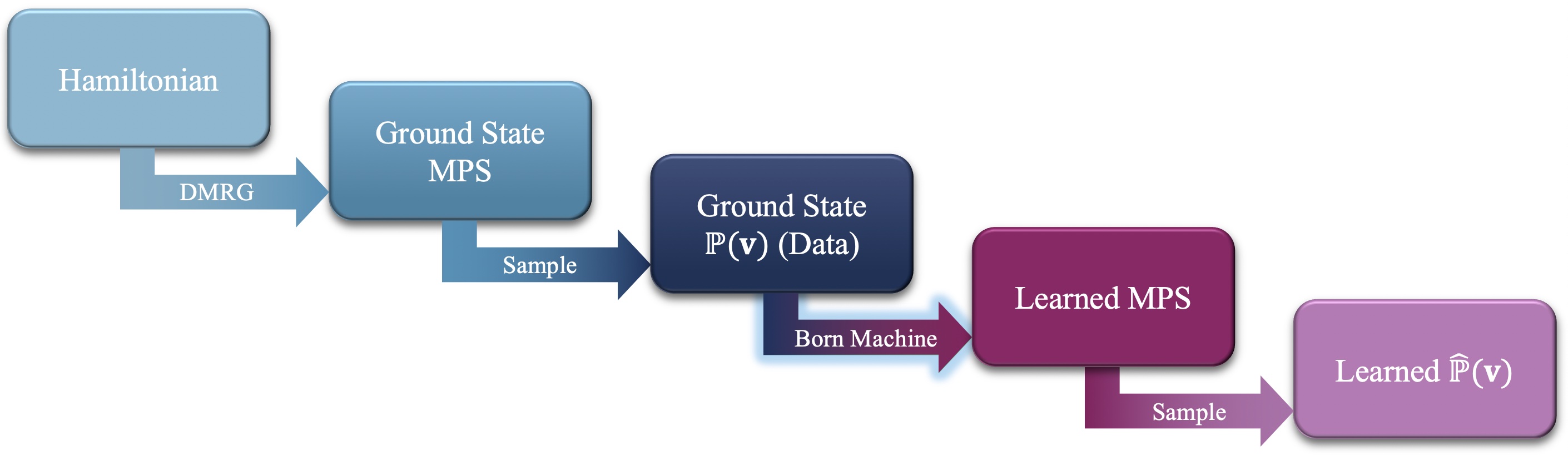}
    \caption{General procedure for generating training data for learning and generating new results from the trained model.}
    \label{fig:flow_chart}
\end{figure}

A necessary ingredient to any machine learning algorithm -- quantum or classical -- is data. As a proof of concept, in this work we assembled training data by sampling from DMRG simulations (see Fig. \ref{fig:flow_chart}). DMRG has proven to be an effective tool for accurately calculating ground state properties of quantum spin chains, including quantum correlations \cite{Misha2}. A variety of points across the phase diagram were selected for study: the Ising critical point ($\gamma = 1, h = 1$), a point from the ordered phase ($\gamma = 1.5, h = 0.5$), a point from the disordered phase ($\gamma = 2, h = 2$), and a point from the oscillatory region ($\gamma = 0.5, h = 0.5$). DMRG was used to calculate the ground state at these points of interest for chain lengths of $N = $ 13, 16, and 19 atoms. Both open and periodic boundary conditions were considered for comparison (OBC and PBC, respectively -- note that the first sum in Eq. (\ref{H_XY}) runs from $i = 1$ to $N$ for the case of PBC). The MPS output by DMRG can be efficiently sampled in the computational basis according to the probability distribution of the quantum state, effectively simulating data that might be acquired through measurement. 

\begin{figure}[htp]
    \centering
    \begin{subfigure}[b]{0.58\textwidth}
        \centering
        \captionsetup{justification=raggedright,singlelinecheck=false}
        \includegraphics[scale = 0.8]{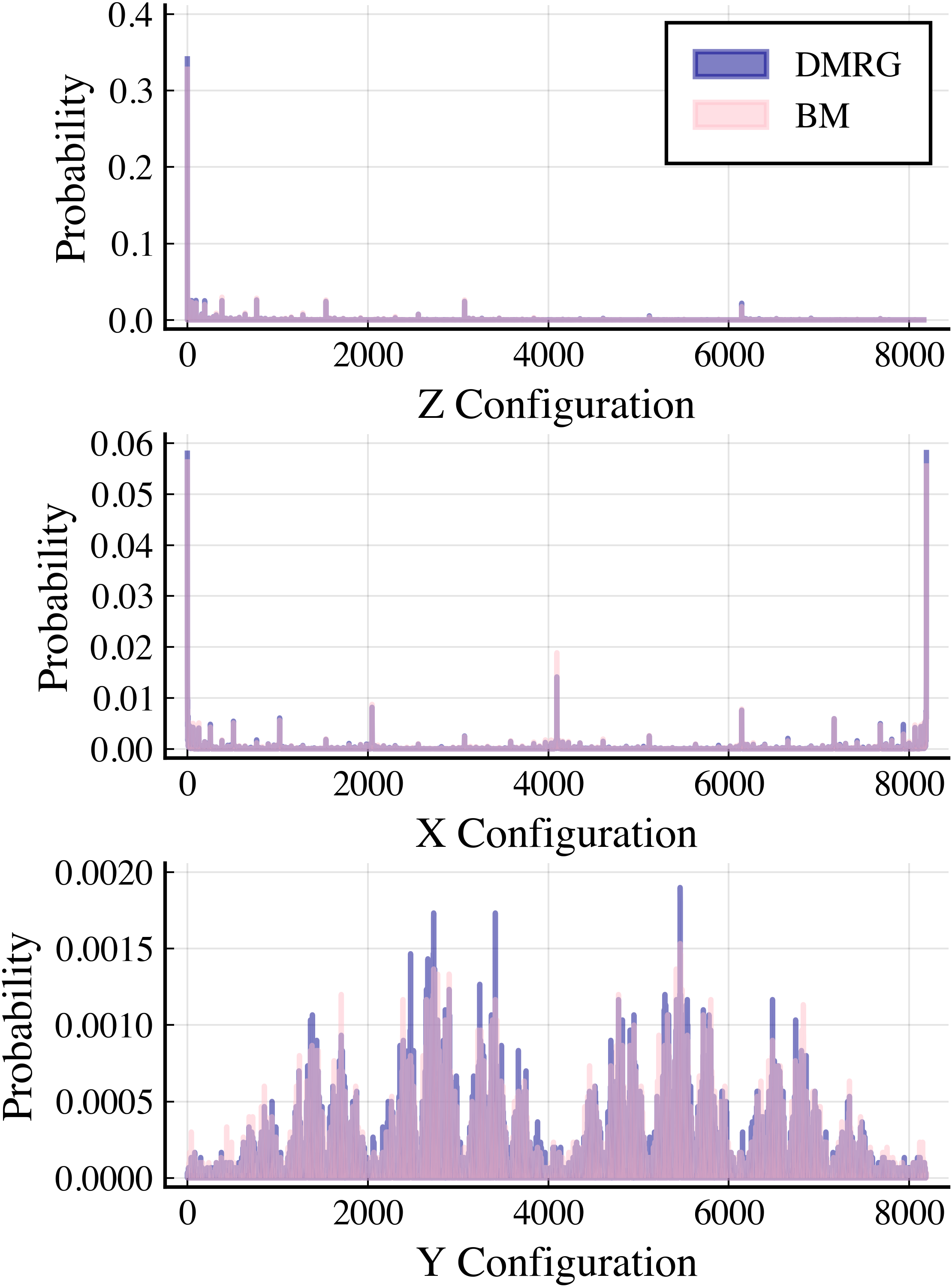}
        \caption{}
        \label{subfig:Prob_Dist}
    \end{subfigure}
     \hfill
    \begin{subfigure}[b]{0.41\textwidth}
        \centering
        \captionsetup{justification=raggedright,singlelinecheck=false}
        \includegraphics[scale = 0.8]{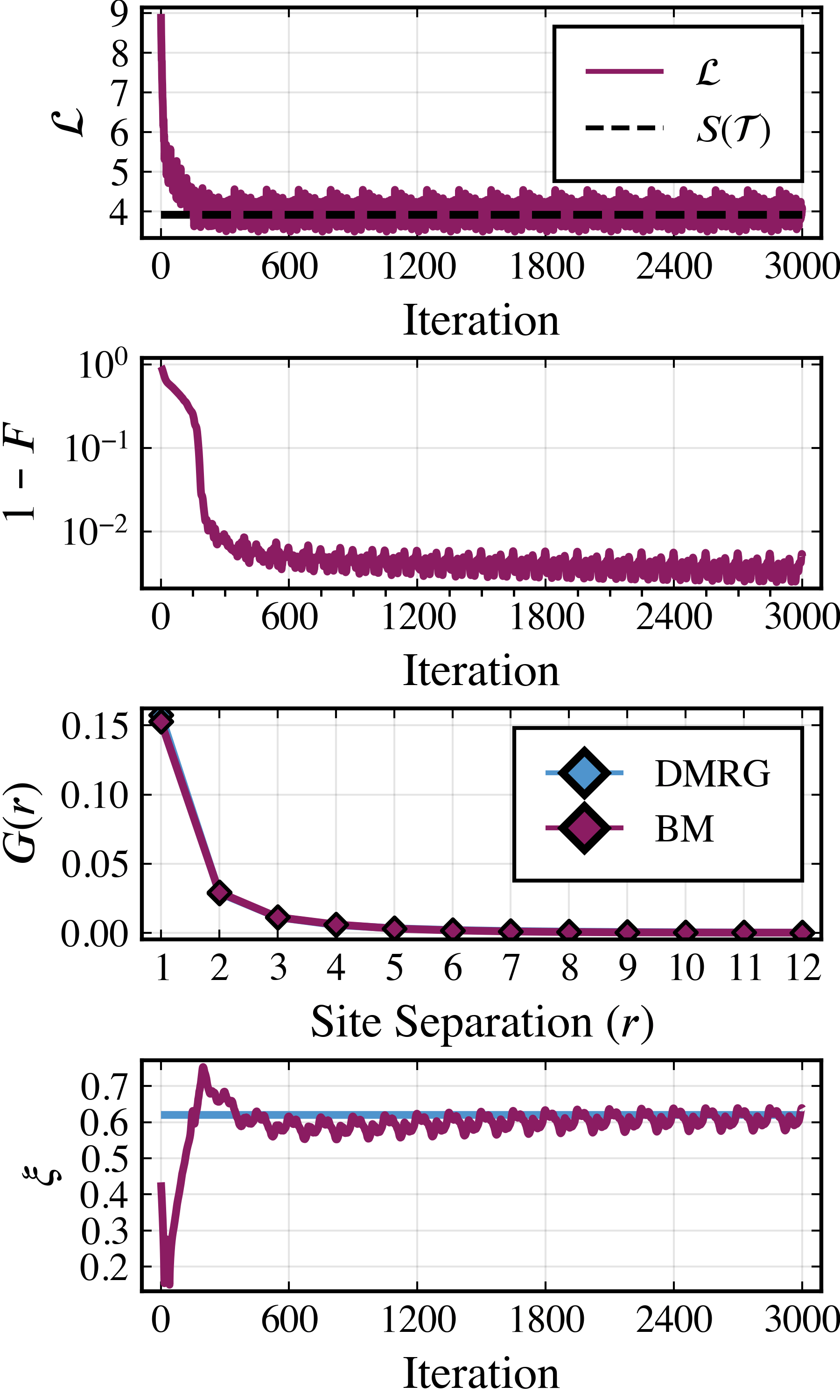}
        \caption{}
        \label{subfig:Loss_and_Corrs}
    \end{subfigure}
    \captionsetup{justification=justified,singlelinecheck=false}
    \caption{(a) Overlaid probability distributions of training data and data generated from the trained MPS at the critical Ising point, for a chain of $N = 13$ atoms with open boundary conditions. The Open Born machine was used for training with $\lvert \mathcal{T} \rvert = 30000$ and bond dimension $D = 4$. (b) Top to bottom: The NLL loss function calculated during training compared with the Shannon entropy of the training data, indicated by $S(\mathcal{T})$. The infidelity $(1-F)$ between the model and the quantum state during training. Only data in the computational basis (that is, the Z configuration) is included in $\mathcal{T}$, indicating the high training power of the Born machine for this model. Correlation function $G(r)$ of the MPS after training, plotted over the correlation function of the DMRG simulations. The correlation length $\xi$ of the model during training plotted alongside the ``target'' correlation length of the DMRG results to highlight the power of Born machine in capturing the long range ordering at the quantum critical point.}
    \label{fig:PDF_Loss_and_Corrs}
\end{figure}

Once the training data is produced, it can be fed into either Born machine to train an MPS model of the state. We used PyTorch's Adam optimizer to train the model with 200 samples per batch and 20 epochs \cite{adam}.
The final MPS can be sampled efficiently, generating new data that can be compared to the original training data \cite{Wang_MPS,Najafi_GHZ2}. Overlapping probability distributions of the training data and data sampled from the trained model (Fig. \ref{subfig:Prob_Dist}) provide visual confirmation that the model can produce data that mimics state measurement. Note that in this figure, data from the $x$- and $y$-bases are included, although only $z$-basis data was used for training. There is high overlap in each basis, indicating that the full quantum state has been learned. It is not generally true that a single-basis Born machine can learn the full quantum state. It has been shown that the MPS based Born machine utilizing gradient based schemes for optimization fails to learn the parity data obtained from GHZ state, which is considered to be nonlocal quantum data \cite{Najafi_GHZ2}. We will further discuss this in Section \ref{4.discussion}.

To further quantify how well the trained MPS captures the quantum state, we examined the loss function during training as well as the quantum fidelity between our model and the quantum state of the system $F = \lvert \langle \Psi| \hat{\Psi}\rangle \rvert ^2$, which can be calculated exactly by taking the inner product of the trained MPS model and the MPS output of DMRG simulation. This would be a daunting task if measurement data was our only resource, as full state tomography would be required to obtain the quantum state. We include the quantum fidelity here in order to emphasize the power of the Born machine. In addition to these metrics, we investigated how well the Born machine is able to capture an observable -- in particular, the correlation function $G(r)$. $G(r)$ is known to provide insightful information about the underlying phase of matter. It is defined as: 
\begin{equation}
    G(r) = \sum_{i}\frac{\langle \sigma^z_{i}\sigma^z_{i+r} \rangle - \langle \sigma^z_{i} \rangle \langle \sigma^z_{i+r} \rangle}{N_{p}}
\end{equation}
Here, $\sigma_z^{i}$ denotes the Pauli $z$ operator acting on the $i^{th}$ site, and the normalization factor $N_p$ is the number of distinct pairs of atoms in the chain separated by $r$. One expects that the correlation function will decay exponentially with distance, e.g: $G(r)=A\, e^{-r/\xi}$, where $\xi$ represents the correlation length. $\xi$ is known to provide information about the underlying quantum phase; for example, due to the long range ordering close to a critical point, the correlation function exhibits a power law decay and the corresponding correlation length is expected to be larger.

\begin{figure}[htp]
    \centering
    \begin{subfigure}[b]{0.32\textwidth}
        \centering
        \captionsetup{justification=raggedright,singlelinecheck=false}
        \includegraphics[scale = 0.8]{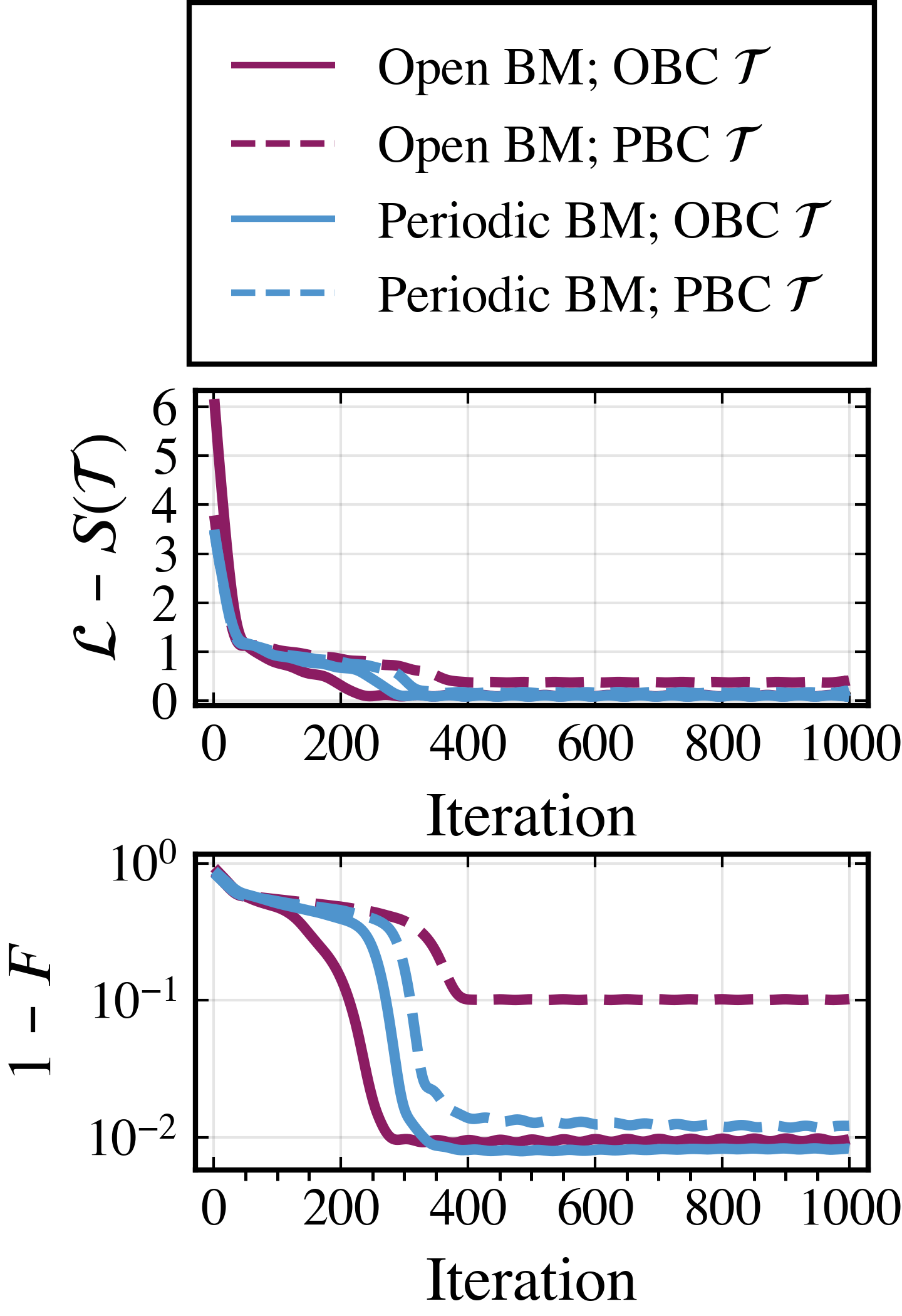}
        \caption{}
        \label{subfig:loss_comp1}
    \end{subfigure}
    \hfill
    \begin{subfigure}[b]{0.32\textwidth}
        \centering
        \captionsetup{justification=raggedright,singlelinecheck=false}
        \includegraphics[scale = 0.8]{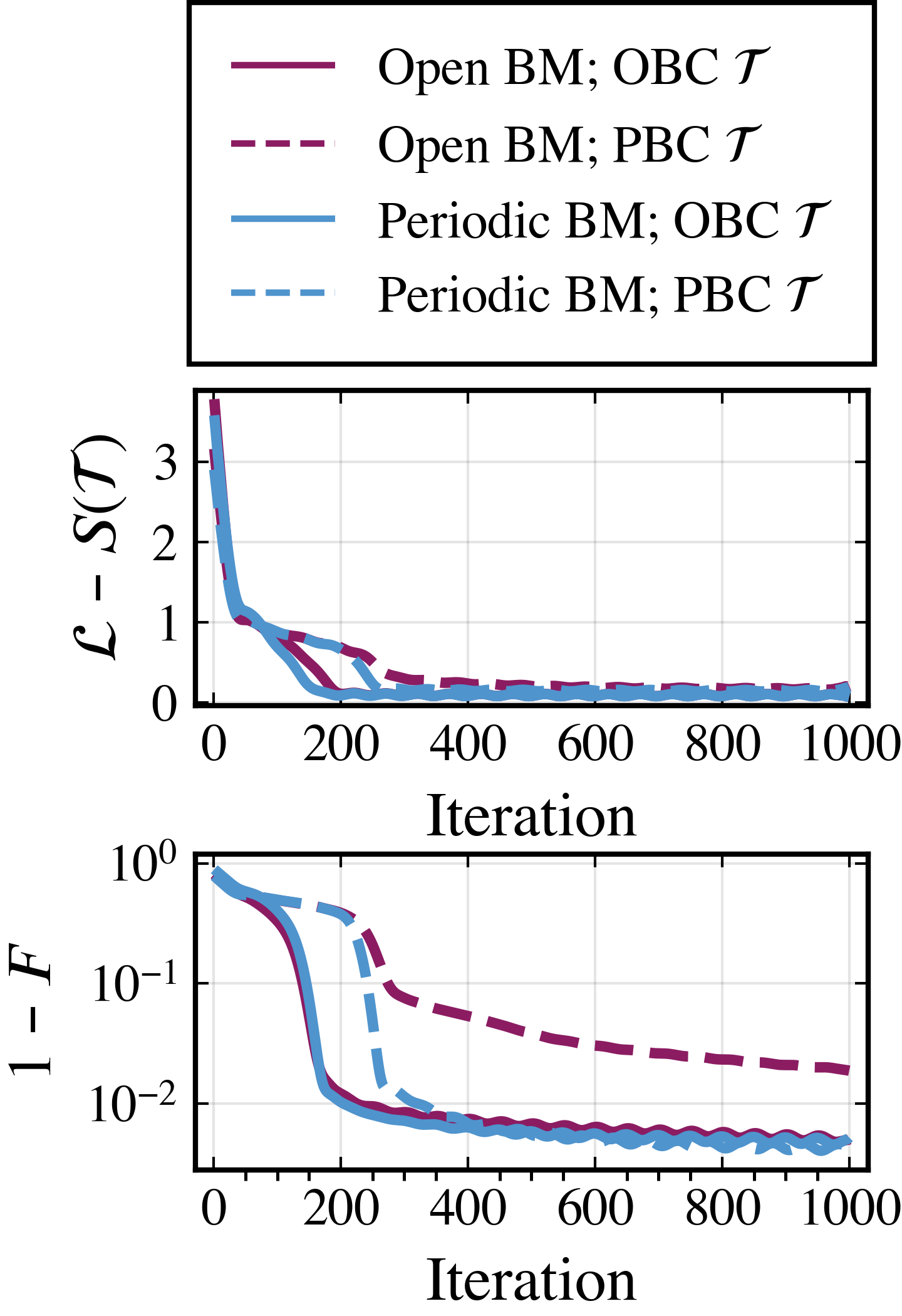}
        \caption{}
        \label{subfig:loss_comp2}
    \end{subfigure}
    \hfill
    \begin{subfigure}[b]{0.32\textwidth}
        \centering
        \captionsetup{justification=raggedright,singlelinecheck=false}
        \includegraphics[scale = 0.8]{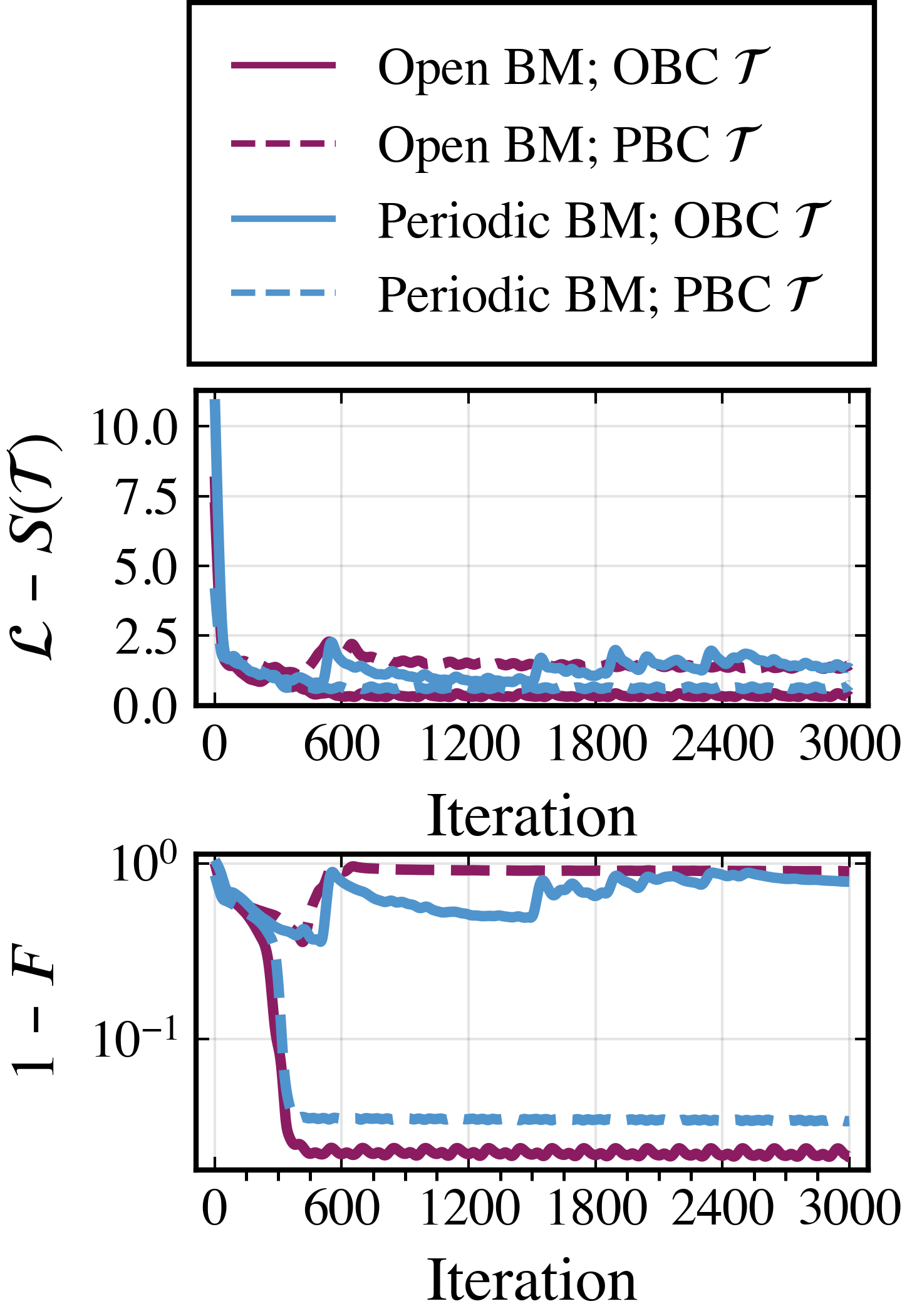}
        \caption{}
        \label{subfig:loss_comp3}
    \end{subfigure}
    \caption{(a) Loss and fidelity during training of the Ising critical point for a chain of 13 atoms with $\lvert \mathcal{T} \rvert =10000$ and bond dimension $D = 2$. (b) Same as (a), with bond dimension increased to $D = 4$. (c) Same as (a), with chain increased to 19 atoms and training data increased to $\lvert \mathcal{T} \rvert =30000$. The loss and fidelity data presented here has been smoothed for clarity of comparison.}
    \label{fig:Loss_Comparison}
\end{figure}

\begin{figure}[htp]
    \centering
    \begin{subfigure}[b]{0.49\textwidth}
        \centering
        \captionsetup{justification=raggedright,singlelinecheck=false}
        \includegraphics[scale = 0.8]{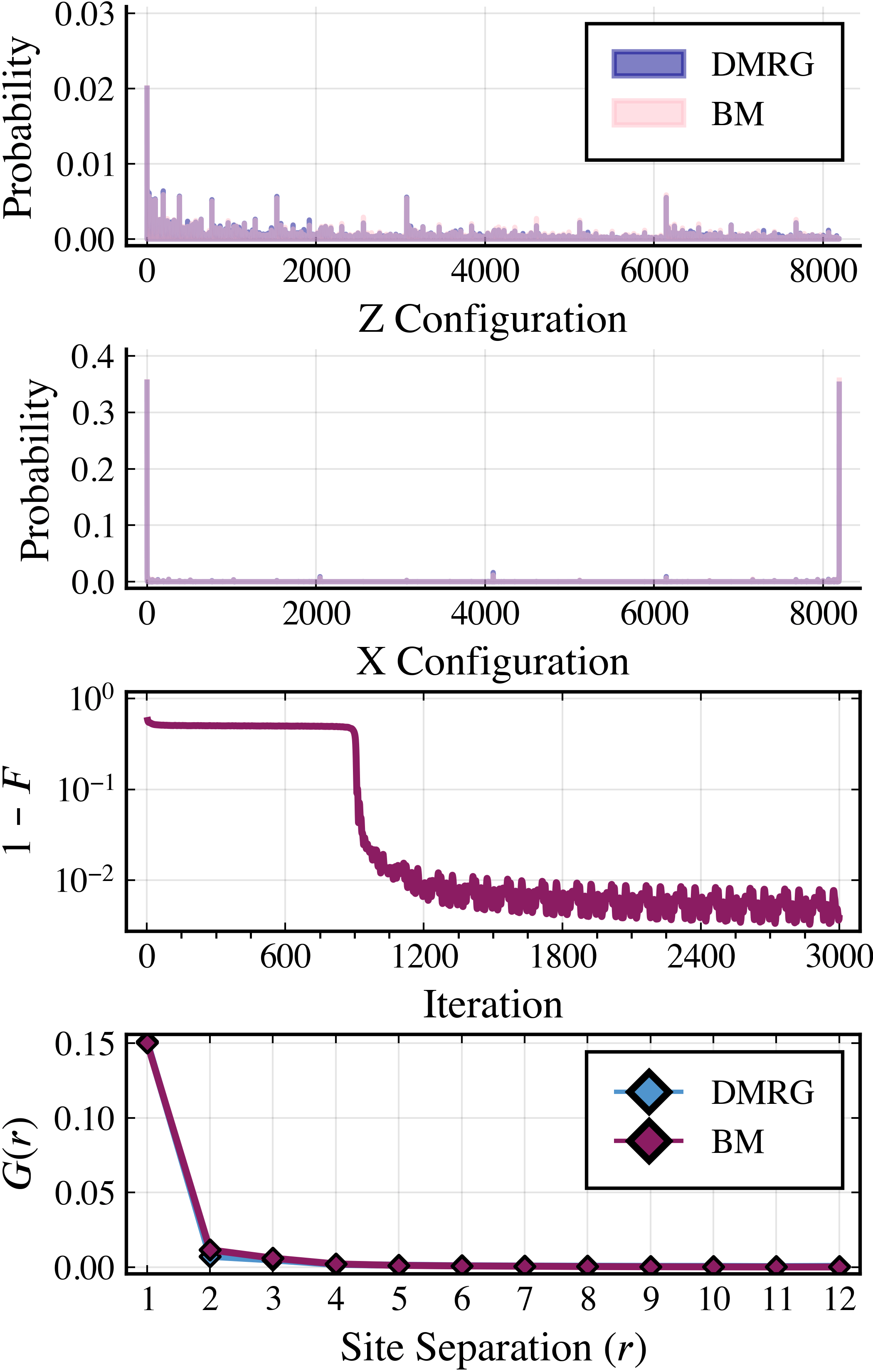}
        \caption{}
        \label{subfig:ordered}
    \end{subfigure}
     \hfill
    \begin{subfigure}[b]{0.49\textwidth}
        \centering
        \captionsetup{justification=raggedright,singlelinecheck=false}
        \includegraphics[scale = 0.8]{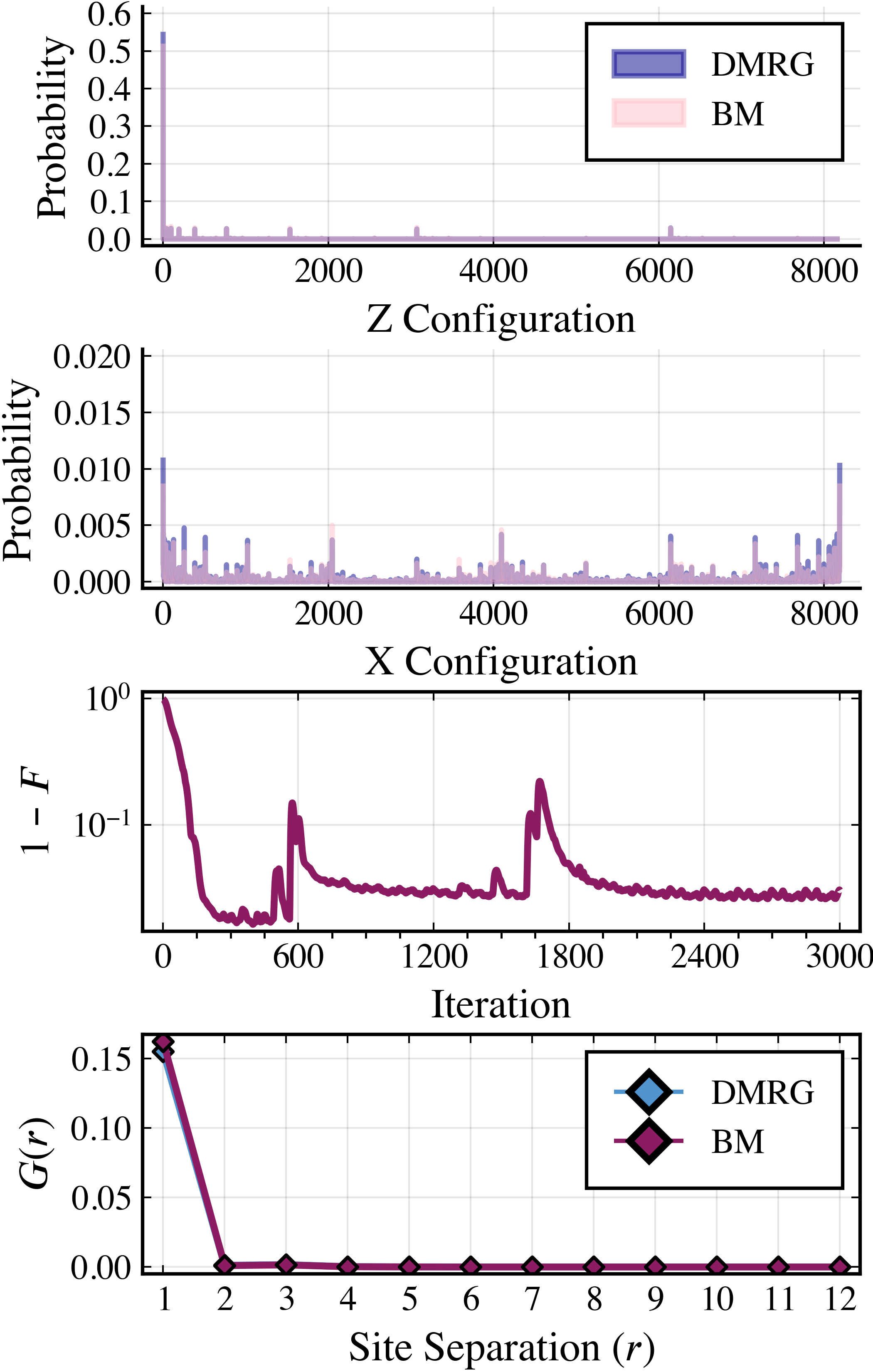}
        \caption{}
        \label{subfig:disordered}
    \end{subfigure}
    \caption{(a) Top two panels: overlaid probability distributions of training data and data generated from the trained MPS at the point $\gamma = 1.5, h = 0.5$ from the ordered phase, for a chain of $N = 13$ atoms with open boundary conditions. The Open Born machine was used for training with $\lvert \mathcal{T} \rvert = 30000$ and bond dimension $D = 4$. The infidelity $(1-F)$ between the model and the quantum state during training. Correlation function $G(r)$ of the MPS after training, plotted over the correlation function of the DMRG simulations. (b) Top two panels: overlaid probability distributions of training data and data generated from the trained MPS at the point $\gamma = 2, h = 2$ from the disordered phase, for a chain of $N = 13$ atoms with open boundary conditions. The Open Born machine was used for training with $\lvert \mathcal{T} \rvert = 30000$ and bond dimension $D = 4$. The infidelity $(1-F)$ between the model and the quantum state during training. Correlation function $G(r)$ of the MPS after training, plotted over the correlation function of the DMRG simulations.}
    \label{fig:Ordered_and_Disordered}
\end{figure}

We first examine the power of the Born machine in learning the quantum critical point. We feed data from the ground state of the transverse field Ising quantum spin chain with length $N = 13$ at the critical point ($\gamma=1,h=1$), simulated by DMRG.  It has been shown that due to the long range ordering at a critical point, both the Boltzmann machine and Born Machine based on projected entangled pair state (PEPS) have failed to learn the 2-dimensional Ising model \cite{Carrasquilla2017, Azizi_Ising}. To our surprise, the MPS-based Born Machine was able to successfully capture the state at the  critical point. In Fig. \ref{subfig:Loss_and_Corrs} we present the training power of the MPS-based Born machine for a variety of quantities, including the correlation function and correlation length.  The loss approaches the Shannon entropy of the training data, indicating that the Born machine was able to fully learn and memorize the data. This is perhaps more evident from the fidelity, which approaches $>99\%$ at the end of the training period. Furthermore, the local Pauli operators ($\sigma_x$ being the order parameter) were also successfully learned as the generated data has good overlap with the original data. We further demonstrate the power of the Born machine in learning the correlation function. The bottom left panels of Fig. \ref{subfig:Loss_and_Corrs} indicate that the MPS Born machine is able to capture the long range correlations at the quantum critical point. Additional results for different points across the phase diagram are provided in Fig. \ref{fig:Ordered_and_Disordered}.

While it is more natural to consider an open boundary condition, a periodic boundary condition allows translational invariance in the physical system, leading to an advantage in the mathematical description and a solution for quantum spin chains. Consequently, the boundary condition imposes different behaviour in physical quantities as well. In order to better address the learnability of the boundary condition and uncover any advantage in using one Born machine architecture over the other, here we present all combinations of boundary conditions between training data and Born machine architecture. 
\begin{figure}[htpp]
    \centering
    \begin{subfigure}[b]{0.54\textwidth}
        \centering
        \captionsetup{justification=raggedright,singlelinecheck=false}
        \includegraphics[scale = 0.8]{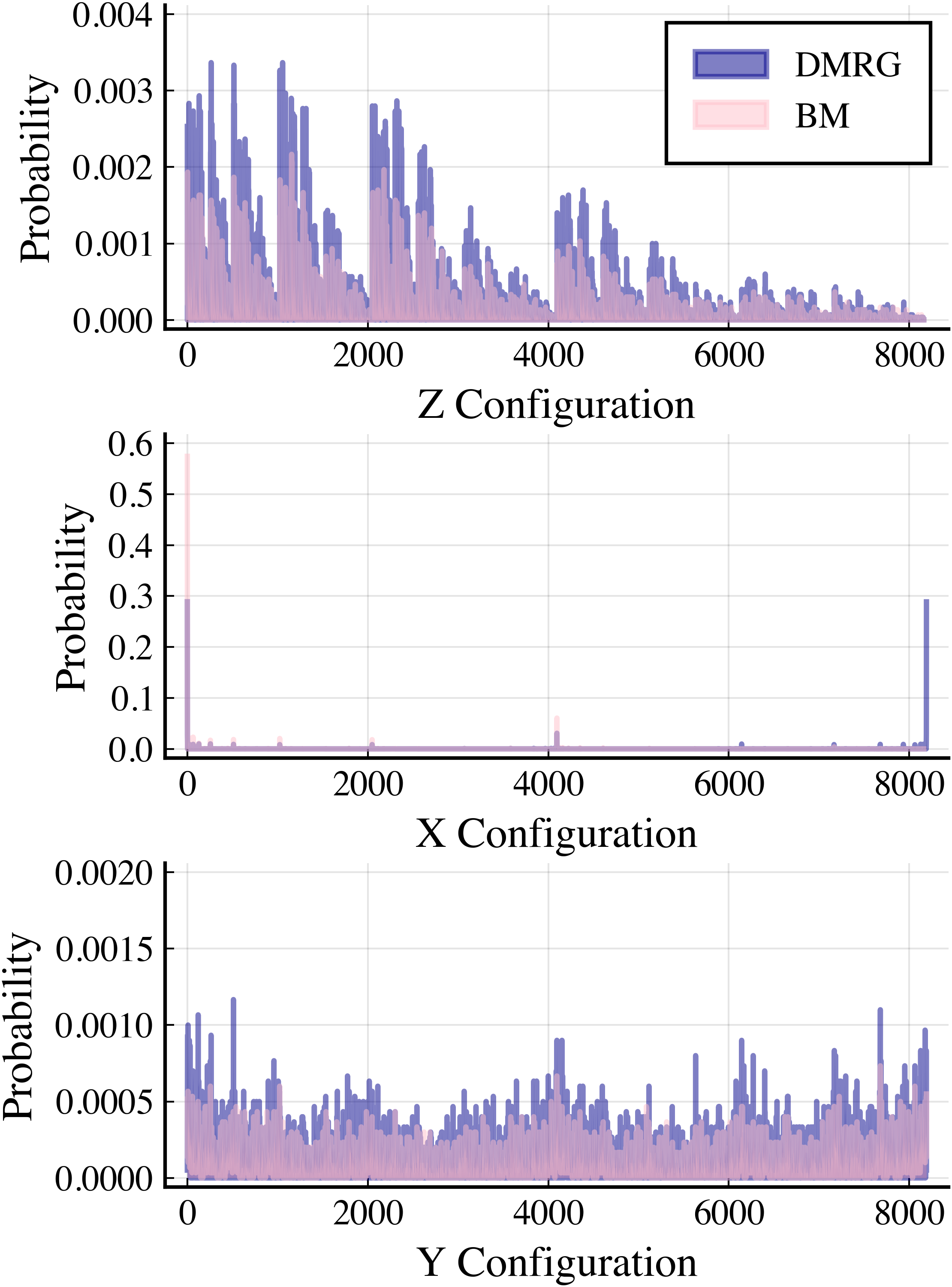}
        \caption{}
        \label{subfig:Prob_Dist2}
    \end{subfigure}
    \hfill
    \begin{subfigure}[b]{0.4\textwidth}
        \centering
        \captionsetup{justification=raggedright,singlelinecheck=false}
        \includegraphics[scale = 0.8]{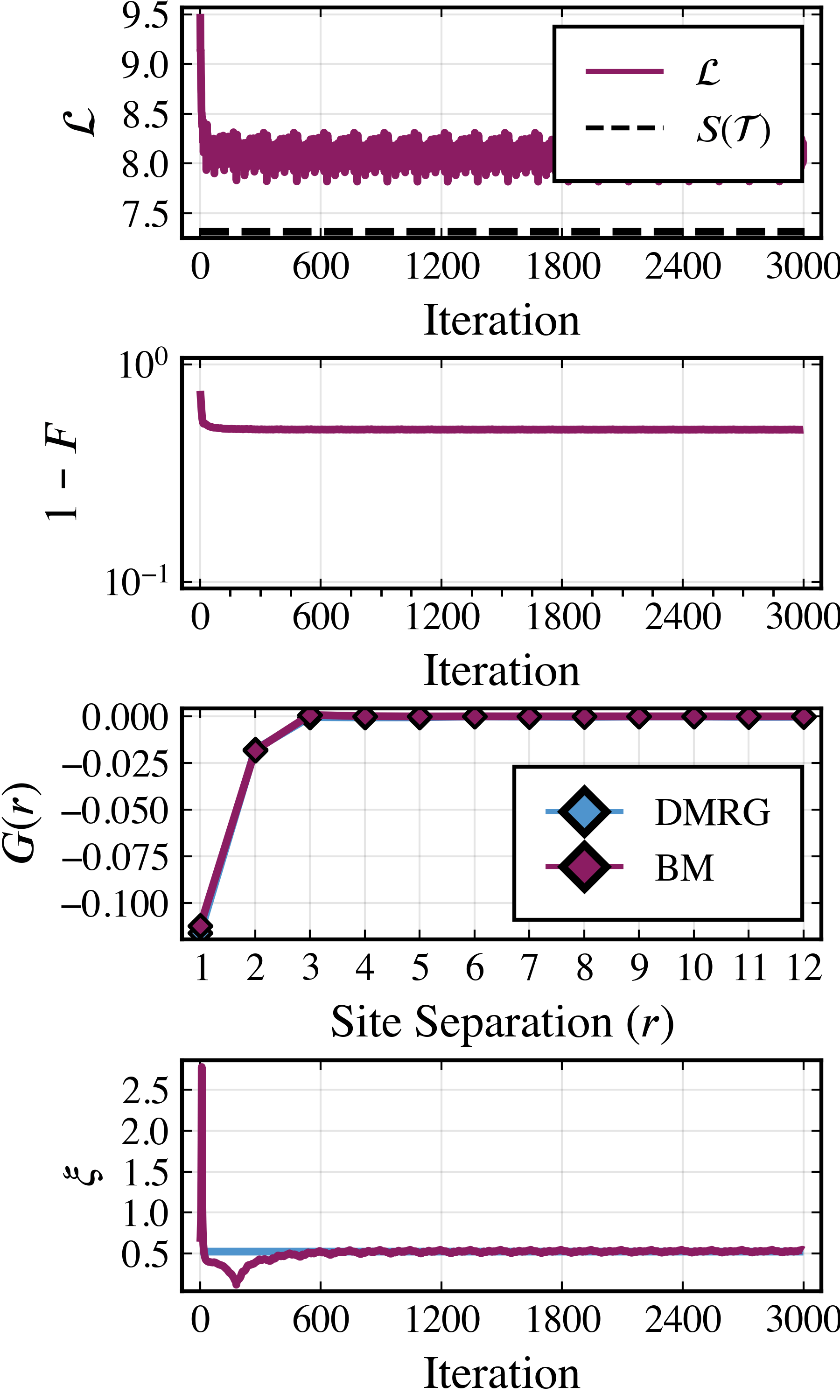}
        \caption{}
        \label{subfig:Loss_and_Corrs2}
    \end{subfigure}
    \caption{(a) Overlaid probability distributions of training data and data sampled from the trained MPS at the point $h = 0.5, \gamma = 0.5$ for a chain of 13 atoms with open boundary conditions. The Open Born machine was used for training with $\lvert \mathcal{T} \rvert = 30000$ and bond dimension $D = 4$. (b) Top to bottom: Loss of the model during training with the Shannon entropy of the training data $S(T)$. Fidelity between the model and the quantum state during training. Correlation function $G(r)$ of the model after training, plotted over the correlation function of the DMRG results. The correlation length $\xi$ of the model during training, plotted alongside the ``target'' correlation length of the DMRG results.}
    \label{fig:PDF_Loss_and_Corrs2}
\end{figure}
\begin{table}[]
     \centering
     {\renewcommand{\arraystretch}{1.15}\begin{tabular}{|c|c|c|c|c|}
          \hline
          Model & Data & $D$ & $\Omega$ & $F$ \\
          \hline \hline
          \multirow{6}{*}{Open}& \multirow{6}{*}{Open} & 2 & 96 & 0.9888 \\
           &  & 3 & 210 & 0.9924 \\
           &  & 4 & 368 & 0.9936 \\
           &  & 6 & 816 & 0.9944 \\
           &  & 8 & 1440 & 0.9944 \\
           &  & 10 & 2240 & 0.9946 \\
          \hline
          \multirow{6}{*}{Open}& \multirow{6}{*}{Periodic} & 2 & 96 & 0.8958 \\
           &  & 3 & 210 & 0.9491 \\
           &  & 4 & 368 & 0.9808 \\
           &  & 6 & 816 & 0.9909 \\
           &  & 8 & 1440 & 0.9908 \\
           &  & 10 & 2240 & 0.9910 \\
          \hline
          \multirow{6}{*}{Periodic}& \multirow{6}{*}{Open} & 2 & 104 & 0.9922 \\
           &  & 3 & 234 & 0.9928 \\
           &  & 4 & 416 & 0.9935 \\
           &  & 6 & 936 & 0.9949 \\
           &  & 8 & 1664 & 0.9944 \\
           &  & 10 & 2600 & 0.9935 \\
          \hline
          \multirow{6}{*}{Periodic}& \multirow{6}{*}{Periodic} & 2 & 104 & 0.9882 \\
           &  & 3 & 234 & 0.9948 \\
           &  & 4 & 416 & 0.9961 \\
           &  & 6 & 936 & 0.9964 \\
           &  & 8 & 1664 & 0.9966 \\
           &  & 10 & 2600 & 0.9964 \\
          \hline
     \end{tabular}}
     \vspace{2mm}
     \caption{Fidelity after training the Ising critical point for a chain of 13 atoms with $|\mathcal{T}| = 10000$. Various combinations of model architecture, data boundary condition, and bond dimension are provided. Bond dimension and model architecture fix the total number of model parameters, $\Omega$.}
     \label{tab:fids}
 \end{table}
Fig. \ref{fig:Loss_Comparison} provides a few examples of learning the Ising critical point for different bond dimensions $D$ and atom chain lengths $N$. Our computations indicate that in general, if enough training data is supplied and a large enough bond dimension is selected, each Born machine architecture is successful in learning from either OBC or PBC data. Perhaps unsurprisingly, in Fig. \ref{subfig:loss_comp1}, for a spin chain of $N=13$ and with only 10000 samples, we see that the Periodic Born machine performs considerably better than the Open Born machine when learning from PBC data. One might attribute this result to the larger expressibility of the Periodic Born machine due to its increased number of model parameters. However, this explanation seems insufficient when one considers that the infidelity was decreased by an order of magnitude with the addition of only eight ($= 2i(D^2 - D)$) parameters. It is the strategic location of these parameters in the architecture as a whole, greatly expanding the connectivity and entanglement-encoding potential of the MPS, that contributes to the success of the Periodic Born machine. This is further explored in Table \ref{tab:fids}, which lists the final fidelities for different combinations of model architecture and data boundary condition for a variety of bond dimensions $D$. The exact number of model parameters, $\Omega$, is provided for clarity. Notice the success of the Periodic Born machine in learning from PBC data; the quantum fidelity reaches $> 99\%$ for $D = 3$ ($\Omega = 234$). In order for the Open Born machine to reach this level of quantum fidelity, a bond dimension of at least 6 is required ($\Omega = 816$). Thus, one can attribute the success of the Periodic Born machine to the architecture of the model -- the physical placement of the model parameters -- rather than their total number. In all situations, the fidelity reaches $> 99\%$ well before $D = 10$ with little improvement beyond that threshold. 

In Fig. \ref{subfig:loss_comp2}, the bond dimension was increased from $2$ to $4$. The Open Born machine was more successful in learning PBC data with more parameters in the model, though it was still out-performed by the Periodic Born machine. Finally, in Fig. \ref{subfig:loss_comp3}, we consider the scenario where there is limited data available (relative to the size of the Hilbert space, as $30000 \ll 2^N$ for $N = 19$) and a small bond dimension has been selected for the model. In this situation, interestingly, the advantage of matching the boundary conditions of the model to those of the data is apparent: for both OBC and PBC data, the opposite model is unable to minimize the loss function and learn the state, while the ``matching'' model can learn the state with fidelities $> 90\%$ -- despite the few number of parameters and small size of the training data. 

\section{Discussion and future directions}
\label{4.discussion}
The transverse field XY Hamiltonian is a simple model manifesting two distinct quantum phase transitions. The exact mathematical solution of this model has provided a powerful tool to investigate a wide variety of observables, making the XY model very valuable in understanding many aspects of quantum phase transitions and in benchmarking. In this paper, we investigated the power of the MPS-based Born machine in learning across the quantum phase transition from disorder to order (see Fig. \ref{fig:Ordered_and_Disordered}), as well as at critical point, (see Fig. \ref{fig:PDF_Loss_and_Corrs}). Surprisingly, our simple Born machine was able to learn the quantum states across different phases and even at critical point despite the long range ordering. We also investigated the learnabilty of the periodic and open Born machines when different boundary conditions are imposed on the training data. Our numerical results indicate that when given enough sampling data and a large enough bond dimension, the Periodic Born machine has a considerably better performance in learning from training data with a periodic boundary condition, and comparable performance in learning from training data with an open boundary condition. This can partially be attributed to the extra parameters provided by the extra bond dimension, but is largely due to the enhanced connectivity that the strategically placed extra bond dimension provides. However, for limited amounts of training data and a small bond dimension, the Born machine becomes sensitive to the original boundary imposed on the quantum data, and the matching Born machine performs better than the opposite architecture, as shown in Fig. \ref{subfig:loss_comp3}.

Finally, although many points on the phase diagram can be learned to a high degree of accuracy by the Born machine (as evidenced by Fig. \ref{fig:PDF_Loss_and_Corrs}), this is not true of every point. In particular, the Born machine is less successful in learning points from the oscillatory region of the phase diagram, where the correlation function shows oscillatory behavior. This region is highlighted in blue in Fig. \ref{Phase_XY}. See Fig. \ref{fig:PDF_Loss_and_Corrs2} for an example of learning the point $h = 0.5, \gamma = 0.5$. The $z$-basis probability distribution associated with this point is spread across most of the Hilbert space, resulting in a large value for the Shannon entropy. This is an obstacle to learning -- the loss does not reach the global minimum, and although some aspects of the state can be captured by the model (such as the correlation function and correlation length), the full quantum state has not been learned faithfully. It is possible to learn such states by incorporating training data from a different basis with a lower Shannon entropy (the $x$-basis, for this particular phase of the system) and introducing complex parameters to increase the expressibility of the model, but this discussion will be left for a future study. 


\section*{Acknowledgments and disclosure of funding}
A.M.G acknowledges support from the NSF through the Graduate Research Fellowships Program, as well as support through the Theodore H. Ashford Fellowships in the Sciences. S.F.Y would like to acknowledge funding by the NSF and DOE.

\bibliographystyle{ieeetr}
\bibliography{main.bib}

\begin{thebibliography}{10}

\bibitem{Misha1}
A.~Omran, H.~Levine, A.~Keesling, G.~Semeghini, T.~T. Wang, S.~Ebadi,
  H.~Bernien, A.~S. Zibrov, H.~Pichler, S.~Choi, J.~Cui, M.~Rossignolo,
  P.~Rembold, S.~Montangero, T.~Calarco, M.~Endres, M.~Greiner, V.~Vuletić,
  and M.~D. Lukin, ``Generation and manipulation of {Schrödinger} cat states
  in {Rydberg} atom arrays,'' {\em Science}, vol.~365, pp.~570--574, Aug. 2019.

\bibitem{Ebadi2021}
S.~Ebadi, T.~T. Wang, H.~Levine, A.~Keesling, G.~Semeghini, A.~Omran,
  D.~Bluvstein, R.~Samajdar, H.~Pichler, W.~W. Ho, S.~Choi, S.~Sachdev,
  M.~Greiner, V.~Vuleti{\'{c}}, and M.~D. Lukin, ``Quantum phases of matter on
  a 256-atom programmable quantum simulator,'' {\em Nature}, vol.~595,
  pp.~227--232, Jul 2021.

\bibitem{Iris_QCNN}
I.~Cong, S.~Choi, and M.~D. Lukin, ``Quantum convolutional neural networks,''
  {\em Nature Physics}, vol.~15, pp.~1273--1278, Dec. 2019.
\newblock arXiv: 1810.03787.

\bibitem{huang_QML_2021}
H.-Y. Huang, R.~Kueng, G.~Torlai, V.~V. Albert, and J.~Preskill, ``Provably
  efficient machine learning for quantum many-body problems,'' {\em
  arXiv:2106.12627 [quant-ph]}, July 2021.
\newblock arXiv: 2106.12627.

\bibitem{Polkovnikov}
A.~Polkovnikov, K.~Sengupta, A.~Silva, and M.~Vengalattore,
  ``\textit{{Colloquium}} : {Nonequilibrium} dynamics of closed interacting
  quantum systems,'' {\em Reviews of Modern Physics}, vol.~83, pp.~863--883,
  Aug. 2011.

\bibitem{Misha2}
A.~Keesling, A.~Omran, H.~Levine, H.~Bernien, H.~Pichler, S.~Choi, R.~Samajdar,
  S.~Schwartz, P.~Silvi, S.~Sachdev, P.~Zoller, M.~Endres, M.~Greiner,
  V.~Vuletić, and M.~D. Lukin, ``Quantum {Kibble}–{Zurek} mechanism and
  critical dynamics on a programmable {Rydberg} simulator,'' {\em Nature},
  vol.~568, pp.~207--211, Apr. 2019.

\bibitem{Franchini}
F.~Franchini, ``An introduction to integrable techniques for one-dimensional
  quantum systems,'' {\em arXiv:1609.02100 [cond-mat, physics:hep-th]},
  vol.~940, 2017.
\newblock arXiv: 1609.02100.

\bibitem{Najafi1}
K.~Najafi and M.~A. Rajabpour, ``Formation probabilities and {Shannon}
  information and their time evolution after quantum quench in the
  transverse-field {XY} chain,'' {\em Physical Review B}, vol.~93, p.~125139,
  Mar. 2016.

\bibitem{Najafi2}
K.~Najafi and M.~A. Rajabpour, ``Area law and universality in the statistics of
  subsystem energy,'' {\em Physical Review B}, vol.~99, p.~075152, Feb. 2019.

\bibitem{Najafi3}
K.~Najafi, M.~A. Rajabpour, and J.~Viti, ``{Return amplitude after a quantum
  quench in the XY chain},'' {\em J. Stat. Mech.}, vol.~1908, p.~083102, 2019.

\bibitem{Najafi4}
K.~Najafi and M.~A. Rajabpour, ``{Entanglement entropy after selective
  measurements in quantum chains},'' {\em JHEP}, vol.~12, p.~124, 2016.

\bibitem{DMRG1}
S.~R. White, ``Density matrix formulation for quantum renormalization groups,''
  {\em Phys. Rev. Lett.}, vol.~69, pp.~2863--2866, Nov 1992.

\bibitem{DMRG2}
U.~Schollwöck, ``The density-matrix renormalization group in the age of matrix
  product states,'' {\em Annals of Physics}, vol.~326, pp.~96--192, Jan. 2011.

\bibitem{Wang_MPS}
Z.-Y. Han, J.~Wang, H.~Fan, L.~Wang, and P.~Zhang, ``Unsupervised generative
  modeling using matrix product states,'' {\em Physical Review X}, vol.~8,
  p.~031012, July 2018.

\bibitem{Wang_TTN}
S.~Cheng, L.~Wang, T.~Xiang, and P.~Zhang, ``Tree tensor networks for
  generative modeling,'' {\em Physical Review B}, vol.~99, p.~155131, Apr.
  2019.

\bibitem{itensor}
M.~Fishman, S.~R. White, and E.~M. Stoudenmire, ``The {ITensor} {Software}
  {Library} for {Tensor} {Network} {Calculations},'' {\em arXiv:2007.14822
  [cond-mat, physics:physics]}, July 2020.
\newblock arXiv: 2007.14822.

\bibitem{adam}
D.~P. Kingma and J.~Ba, ``Adam: {A} {Method} for {Stochastic} {Optimization},''
  {\em arXiv:1412.6980 [cs]}, Jan. 2017.
\newblock arXiv: 1412.6980.

\bibitem{Najafi_GHZ2}
K.~Najaﬁ, M.~E. Stoudenmire, S.~Yelin, A.~Azizi, X.~Gao, M.~D. Lukin, and
  M.~Mohseni, ``Limitations of gradient-based {Born} {Machines} over tensor
  networks on learning quantum nonlocality,'' in {\em 34th Conference on Neural
  Information Processing Systems}, First Workshop on Quantum Tensor Networks in
  Machine Learning, 2020.

\bibitem{Carrasquilla2017}
J.~Carrasquilla and R.~G. Melko, ``Machine learning phases of matter,'' {\em
  Nature Physics}, vol.~13, pp.~431--434, May 2017.

\bibitem{Azizi_Ising}
A.~Azizi, K.~Najaﬁ, and M.~Mohseni, ``Learning {Phase} {Transition} in
  {Ising} {Model} with {Tensor}-{Network} {Born} {Machines},'' in {\em 34th
  Conference on Neural Information Processing Systems}, First Workshop on
  Quantum Tensor Networks in Machine Learning, 2020.

\end{thebibliography}

\end{document}